# A simple method to estimate entropy of atmospheric gases from their action

Ivan R. Kennedy, Harold Geering, Michael T. Rose and Angus N. Crossan

SUNFix Laboratory, Faculty of Agriculture and Environment, University of Sydney, NSW 2006, Australia

***Correspondence to:** ivan.kennedy@sydney.edu.au

A convenient model for estimating the total entropy ($\Sigma S_i$) of atmospheric gases based on physical action is proposed. This realistic approach is fully consistent with statistical mechanics, but uses the properties of translational, rotational and vibrational action to partition the entropy. When all sources of action are computed as appropriate non-linear functions, the total input of thermal energy ($\Sigma S_i T$) required to sustain a chemical system at specific temperatures ($T$) and pressures ($p$) can be estimated, yielding results in close agreement with published experimental third law values. Thermodynamic properties of gases including enthalpy, Gibbs energy and Helmholtz energy can be easily calculated from simple molecular and physical properties. We propose that these values for entropy are employed both chemically for reactions and physically for computing atmospheric profiles, the latter based on steady state heat flow equilibrating thermodynamics with gravity. We also predict that this application of action thermodynamics may soon provide superior understanding of reaction rate theory, morphogenesis and emergent or self-organising properties of many natural or evolving systems.

## Introduction

As defined by Clausius (1875), entropy can be considered as measuring the “self reservoir of heat required to raise the temperature of a system of ideal gas molecules to the temperature $T$”. Thus, in agreement with the third law of thermodynamics, at the absolute zero of temperature Kelvin (K) the entropy should also be zero.

Clausius named entropy using the Greek word for ‘in-turning’ or transformation, a dynamic definition that this paper will show is highly apt. As a property of state (pressure, temperature and volume), irrespective of the path a system of molecules has arrived at that condition, entropy is therefore an important feature of atmospheric gases. Indeed, its capacity to explain how thermal radiation may be absorbed and partitioned into different degrees of freedom of motion is key information in explaining the warming potential of greenhouse gases.

The total entropy of an ideal gas molecule can be calculated as the sum of terms,

$$S_{\mathrm{Total}} = S_{\mathrm{t}} + S_{\mathrm{r}} + S_{\mathrm{v}} + S_{\mathrm{e}} + \mathrm{S}_{\mathrm{n}} + \ldots. \qquad (1)$$

where the subscripts refer to translational, rotational, vibrational, electronic and nuclear entropy terms respectively. To estimate the total thermal energy needed to reversibly heat (i.e. without doing other work) a system of gas molecules we need only multiply by the temperature *T*. This thermal energy can be thought of as both kinetic and potential energy contained in the field of the molecular system (Kennedy, 2000; 2001). Furthermore, this thermal energy input is physically required to sustain the physical action of the system (Rose *et al*., 2008).

This paper seeks to place thermodynamics within easy reach even of non-specialists by giving a key role to action, a physical property that is realistically the focus of our interest. Action is related to the vector angular momentum, but is a distinct scalar quantity since it is independent of direction. Planck (1913) defined molecular action as a macroscopic property of the product of the generalised space coordinates (*r*, cm) and the generalised impulse coordinates (*mv*, gcm/sec), defining the microscopic state of the molecule (*mvr*). Like entropy, action is an extensive or cumulative property, but with physical dimensions of the integral of energy with time, or of the instantaneous angular momentum with respect to angular motion; classically, action was considered as the integral of momentum with distance. As a variable property of conservative systems, action has been considered to take stationary values, a result sometimes referred to as the principle of least action. In fact, all these viewpoints of action are equivalent. Note that angular motion is the ratio of circumference to radius and thus is physically dimensionless, although we still measure it in degrees or radians.

In illustration of the utility of action theory, we advance a unified model that we will show is valid for calculating the entropy of atmospheric gases. In the range of ambient temperatures relevant to the Earth's atmosphere, this realistic model gives results for absolute entropy closely consistent with previous experimental data. This action model may help provide an approach to prediction of the rate of global warming based on causal responses to the increasing greenhouse gas content of the atmosphere, rather than statistical correlations. In particular, a direct relationship between thermodynamics and gravity may provide a dynamic view of how the thermal properties of the atmosphere have significance for warming and climate change.

**Methods**

Here we show how the classical formulae for estimating translational, rotational and vibrational entropy using partition functions may be reviewed as physical action. Although entropy data for atmospheric gases are readily available in standard tables, the methods developed here illustrate how easily such data can be manipulated to account for real environmental conditions.

Because of their relevance to the following methods, we state here the partition functions for ideal gases currently used to calculate the entropies for molecular translation, rotation and vibration. These functions are given in all standard texts on modern statistical mechanics, the discipline founded jointly by Ludwig Boltzmann (1896) and J. Willard Gibbs (1902). The factors governing these functions include absolute temperature ($T$ in $^{o}$K), Boltzmann's constant ($k$ = 1.3806x10$^{-23}$ J K$^{-1}$), Planck's quantum of action ($h$ = 6.626x10$^{-34}$ J. sec) and the system volume ($V$). For ease of use and consistency in dimensions, all modelling and calculations have been performed in cgs units, before conversion to SI units where required.

*Translational partition function*

$$Q_t = (2\pi mkT/h^2)^{3/2}V$$

Here $V$ is taken as the system volume occupied by the molecules (Glasstone, 1951).

*Rotational partition function (linear molecule)*

$$Q_r = 8\pi IkT/h^2$$

*Rotational partition function (non-linear molecule)*

$$Q_r = 8\pi^2(8\pi^3 I_A I_B I_C)^{1/2}(kT/h^2)^{3/2}$$

*Vibrational partition function (polyatomic molecules)*

$$Q_{vi} = \Pi_i\,[1 - \exp^{-hvi/kT}]^{-1}$$

$hv_i$

where $\Pi_i$ indicates a product of i functions, for each mode of vibration.

It will be shown that these functions can all be considered as dimensionless statistical measures of relative molecular action, using Planck's quantum $h$ or its reduced form per radian, $\hbar$, as a natural reference unit. Note that there are only three sources of variation in the partition functions – inertial mass, temperature or the root mean square velocity and pressure or density.

***Translational entropy and the Sackur-Tetrode equation***

The Sackur-Tetrode equation was published early in the 20$^{th}$ century (Glasstone, 1951), based on Gibbs' theory of statistical mechanics (1902).

$$S_t = R[\ln(2\pi mkT)^{3/2}V/h^3N + 2.5] \qquad (2)$$

This equation allows calculation of the translational entropy of N molecules of an ideal monatomic gas. This is derived from the relationship from the calculus of total entropy including translational $S_t$ and internal $S_{int}$ parts as follows.

$$S = RT(\partial \ln Q/\partial T)_V + R\ln Q - k\ln N! = S_t + S_{int}$$

$$= RT[(\partial \ln Q_{tr}/\partial T) + (d\ln Q_{int}/dT)] + R[\ln Q_{tr} + \ln Q_{in}] - k\ln N!$$

Here the factor $k$lnN! allows for the inability to distinguish between N identical molecules.

For ideal monatomic gases, no internal entropy (from rotation or vibration) at normal temperatures exists, so the differential of the internal partition function $Q_{int}$ in the above equation can be ignored.

$$S_t = RT[(\partial \ln Q_{tr}/\partial T)] + R[\ln Q_{tr}] - k\ln N!$$

Using a solution from the Schrodinger equation that $Q_t = (2\pi mkT/h^2)^{3/2}V$
and the Stirling approximation for lnN! as effectively NlnN - N we have from Moore (1963).

$$S_t = 3/2R + R\ln[(2\pi mkT/h^2)^{3/2}V] - R\ln(N- 1)$$
$$= R[\ln(2\pi mkT)^{3/2}V/(h^3N) + 2.5]$$

Note the close similarity to the translational partition function. This is a well known result of statistical mechanics, with $h$ being the constant introduced by Planck (1913) as the quantum of action for radiation. Despite their lack of rest mass, all energy quanta possess an action of magnitude $h$ and their energy is given by $h\nu$, where $\nu$ is their frequency.

Here we introduce a revised approach, based on the use of the property of state, action (@= *mrv* (Kennedy, 2000; 201). This allows the establishment of the relative action, a ratio or pure number suitable for logarithmic expression. Some expressions of entropy in text books include isolated terms such as the logarithm of the temperature ln*T*. Strictly, this is invalid, as logarithms can only be taken of pure numbers or dimensionless ratios and never of quantities with physical dimensions.

For each of the three translational degrees of freedom, the translational action $@_t$ can be derived from the kinetic energy, given each has ½*kT* of kinetic energy. For three degrees of freedom we can estimate the action, *mvr* or $I_t\omega_t$, as follows. The three-dimensional kinetic energy $\frac{1}{2}mv^2$ is $3/2kT$ which equals $\frac{1}{2}mr_t^2\omega_t^2$ or $\frac{1}{2}I_t\omega_t^2$, for motion with polar coordinates; here the translational angular velocity $\omega_t$ or $d\Theta/dt$ is given in radians per sec. Thus $I_t\omega_t^2$ is equal to $3kT$ and so $I_t\omega_t$ is equal to $3kTI_t/\omega_t$ or $(3kTI_t)^{1/2}$. The translational action $@_t$ is thus defined as equal to $(3kTI_t)^{1/2}$. It should be noted that $3kT$ is a statistical result from the three-dimensional Maxwell distribution equal to twice the most probable kinetic energy $\frac{1}{2}mv^2$ for the root mean square velocity *v* in three dimensions; moreover, 50% of molecules have speeds greater than the root-mean square velocity, which is about 1.09 times the mean speed of the ideal gas molecules (Brown, 1968, Table 8). In the Maxwell distribution, the most probable velocity is slightly less than either of these speeds.

We can regard the system volume *V* as containing N cubic cells of volume $a^3$, a cell for each gas molecule. Then, for $r_t$ arbitrarily taken as the mean value of the half-distance between the centres of any two nearest neighbour gas molecules, $a^3$ is equal to $(2r_t)^3$ or $8r_t^3$. Considering a mole of gas at standard temperature and 1 atmosphere pressure ($N=6.022169 \times 10^{23}$ molecules in 24465.1 mL at STP), then $r_t$ or $(V/N)^{1/3}/2$ is equal to $1.7188 \times 10^{-7}$ cm.

We can then substitute into the Sackur-Tetrode equation ($S_t = R\ln[8e^{5/2}r_t^3(2\pi mkT)^{3/2}/h^3]$). Taking the $r_t^3$ term inside the brackets, we have $S_t$ equal to $R\ln[8e^{5/2}(2\pi mr_t^2kT)^{3/2}/h^3]$.

Then, taking $mr_t^2kT$ equal to $kTI_t$ and $@_t$ equal to $(3kTI)^{1/2}$

$$S_t = R\ln[8(2\pi/3)^{3/2}e^{5/2}(@_t/h)^3]$$

For $\hbar$ equal to $h/2\pi$, the reduced quantum of action, we have,

$$S_t = R\ln[e^{5/2}(2/3\pi)^{3/2}(@_t/\hbar)^3]$$

If the factor of $(2/3\pi)^{3/2}$ were to be incorporated into the action term, the inertial radius would be decreased to 7.92287x10$^{-8}$ cm rather than 1.7188x10$^{-7}$ cm. However, we simply assume in equation (3) a translational resonance symmetry factor $z_t$ of 10.2297 or $(2.1708)^3$, replacing $(2/3\pi)^{3/2}$.

$$S_t = R\ln[e^{5/2}(@_t/\hbar)^3/z_t] \qquad (3)$$

Given the disorderly nature of the spatial distribution of the translating particles (see Figure 1), the symmetry factor cannot have a precise continuous value, but should be statistically distributed and fluctuating with time for each molecule (see Figure 1). With real gases having varying degrees of interaction or binding, the most probable radius may vary from one gas to another. To maintain consistency with Sackur-Tetrode theory, all calculations in this paper have employed the geometric result of 1.7188x10$^{-7}$ cm for $r_t$ at STP, or 1.6994x10$^{-7}$ cm at the Earth's global average surface temperature of 288.15 K. However, it may be possible to make experimental determinations of the dynamic structure, allowing more accurate estimates of the action.  In any case, the sensitivity to variations is low given its logarithmic nature and any errors will only be a slight displacement in the entropy value. Furthermore, the absolute value of the translational symmetry factor $z_t$ is rarely of importance because in nearly all cases, differences in entropy of free energy are being taken, or the system is isothermal. In such cases, the $z_t$ factor disappears.

As a result, with suitable choices of the radius and translational symmetry ($z_t$), we can write the following concise relationship for translational entropy.

$$S_t = R\ln[e^{5/2}(@_t/\hbar)^3/z_t]$$
$$= 2.5R + 3R\ln(@_t/\hbar) - R\ln z_t$$

***Rotational action and entropy***

From statistical mechanics (see Moore, 1963), the rotational contribution to the molar entropy of a diatomic or linear molecule with two-dimensional inertia is given by

$$S_r = R + R\ln(8\pi^2 kTI_r/\sigma_r h^2), \text{ or } R\ln[e(8\pi^2 kTI_r/\sigma_r h^2)] = R\ln[e(2kTI_r/\hbar)^2/\sigma_r]$$

This is two-dimensional only, because there is no significant inertia around the longitudinal axis of a diatomic or linear molecule like $N_2$ or $CO_2$. The rotational partition function is $8\pi^2 kTI_r/h^2$; clearly, this can also be recast as an action ratio. Here the moment of inertia $I_r$ is given by $(m_1m_2/m_1+m_2)r_r^2$ and $r_r$ is the average bond length, $\sigma_r$ is the rotational resonance symmetry number (e.g. $\sigma_r = 2$ for $O_2$ and $\sigma_r = 1$ for NO). In this equation we can recognize the rotational action of a gas molecule, $@_r$ equal to $(2kTI)^{1/2}$ – derived from the rotational energy equal to $\frac{1}{2}mr_r^2\omega_r^2$ or $I_r\omega_r^2/2$. So $I_r\omega_r^2$ equals $2kT$ and $I_r\omega_r$ is given by $(2kTI)^{1/2}$ equal to $@_r$, by definition. As a result, using similar notation as for translational action and entropy, we have

$$S_r = R\ln[e(@_r/\hbar)^2/\sigma_r] = R + R\ln[(@_r/\hbar)^2/\sigma_r] \quad (4)$$

Thus, just as in the case of translational entropy, the rotational entropy can also be expressed as a variable of action alone, given that the symmetry factor $\sigma_r$ is a constant. It is shown elsewhere that action is a function of volume and temperature (Kennedy, 2001) but volume or pressure changes have little or no effect on rotational entropy as long as the temperature is not too high. All other terms in this equation are constant for a given gas molecule.

For non-linear gas molecules with more than two atoms such as the greenhouse gases, other than the linear carbon dioxide and nitrous oxide, the rotational entropy $S_r$ is given from statistical mechanics as,

$$S_r = R\ln[\{8\pi^2(8\pi^3 I_A I_B I_C)^{1/2}(kT)^{3/2}\}/\sigma_r h^3] + 3/2R,$$

where $I_A$, $I_B$ and $I_C$ correspond to the three principal moments of rotational inertia with respect to three perpendicular axes (see Glasstone, 1951). In terms of action ratios analogous to those used above, we rearrange this equation to read,

$$S_r = R\ln[\pi^{1/2}\{(8\pi^2 kTI_A/h^2)^{1/2}(8\pi^2 kTI_B/h^2)^{1/2}(8\pi^2 kTI_C/h^2)^{1/2}\}/\sigma_r] + 3/2R$$

Recalling our previous definition of rotational action $@_r$ of a diatomic molecule as $(2kTI_r)^{1/2}$ for each inertial axis of a linear molecule with more than one atom, we can express the rotational entropy contribution of a non-linear molecule as,

$$S_r = R\ln[\{\pi^{1/2}(@_A/\hbar)(@_B/\hbar)(@_C/\hbar)\}/\sigma_r] + 3/2R$$

$$= R\ln[\{\pi^{1/2}e^{3/2}(@_A @_B @_C/\hbar^3)\}/\sigma_r] \qquad (5)$$

where $@_A$, $@_B$, and $@_C$ are the three principal rotational actions for non-linear molecules. So once again, it is possible to express changes in entropy as a simple function of action alone, as all other terms in the equation are constant for a given gas molecule. Given that the product of entropy and temperature $ST$ indicates the thermal energy required, there is obviously an exact logarithmic relationship between the total energy required to sustain a system of molecules at a given temperature and the action of each mode of rotation

The rotational symmetry number $\sigma_r$ for polyatomic molecules depends on the point group of the molecule as defined by the Nobel laureate Herzberg (1945), by "A possible combination of symmetry operations that leaves at least one point unchanged is called a point group". This is a term derived from crystallography and the characteristic symmetry number σ of each point group can be shown to be equal to "the number of indistinguishable positions into which the molecule can be turned by simple rigid rotations" (Herzberg, 1950). Table 1 adapted from Herzberg (1945) gives the symmetry number for the more important point groups. Note that methane in the *T* point group has a rotational symmetry of 12, indicating how its quantum field indicated by its rotational entropy is economical for energy in view of its indistinguishable structure regarding its orientation in space. This situation for methane can be contrasted with a similar tetrahedral carbon molecule having only one hydrogen atom in its structure, together with three different halogens such as fluorine, chlorine and bromine. In this case the symmetry is unity (1.0), so that the energy field has a 12-fold lower frequency of encountering an identical structure in action space.

**Table 1**: Symmetry numbers for various point groups

| Point group | Symmetry No. $\sigma_r$ | Point group | Symmetry No. $\sigma_r$ | Point group | Symmetry No. $\sigma_r$ |
|---|---|---|---|---|---|
| | | | | | |
| $C_1$, $C_i$, $C_s$ | 1 | $D_2$, $D_{2d}$, $D_{2h}$ $V_h$ | 4 | $C_\infty$ | 1 |
| $C_2$, $C_{2v}$, $C_{2h}$ | 2 | $D_3$, $D_{3d}$, $D_{3h}$ | 6 | $D_{\infty h}$ | 2 |
| $C_3$, $C_{3v}$, $C_{3h}$ | 3 | $D_4$, $D_{4d}$, $D_{4h}$ | 8 | $T$, $T_d$ | 12 |
| $C_4$, $C_{4v}$, $C_{4h}$ | 4 | $D_6$, $D_{6d}$, $D_{6h}$ | 12 | $O_h$ | 24 |
| $C_6$, $C_{6v}$, $C_{6h}$ | 6 | $S_6$ | 3 | | |

Modified from Herzberg (1945), p. 508

Quite naturally this 3-dimensional relationship we will see later is related to chemical potential rather than kinetics, must be exponential. So more energy per molecule is required to convey information in a molecular field of greater probe-ability,with greater radial separation of identical molecules.

***Vibrational action and entropy***

Vibration in molecules between their atoms occurs at a rate usually much more frequent than rotation, clearly adding to the action of molecules by increasing the length of the trajectory of their atoms in a given time interval. Vibration is characterized by its frequency, which does not change appreciably as more energy is added to a system. On the contrary, the amplitude of vibration increases affecting the mean kinetic and potential energies and the inertial trajectory of the molecular angular motion. In collision processes, a higher non-equilibrium vibrational energy state resulting from absorption of a quantum of energy in the infrared will act to equilibrate with its rotational and translational energies in the micro-wave and radio-wave bands, effectively dissipating vibrational energy into these modes. This concept will be revisited later when we consider greenhouse gases and how they can affect the gravitational distribution of the atmosphere.

Related to the point groups and numbers of atoms in the gas molecule are the number of vibrational modes; also, the question arises of how many of these modes are degenerate, that is some different vibrations have identical frequencies because of similarities in molecular structure. For a non-linear molecule the number of vibrational modes is equal to 3n-6 where n is the number of atoms. For a linear molecule, the number of possible modes is one less or 3n-5. In either case, 3n is the total degrees of freedom for motion. The number 6 in the case of non-linear molecules is the sum of 3 translational and 3 rotational degrees of freedom. The number 5 in the case of linear molecules refers to 3 translational and 2 rotational degrees of freedom.

For both non-degenerate and degenerate modes, the vibrational entropy $S_{vi}$ is given by Glasstone (1951),

$$S_{vi} = R\mathrm{x}/(\mathrm{e}^{\mathrm{x}} - 1) - R\ln(1 - \mathrm{e}^{-\mathrm{x}}), \text{ where } x = hc\nu_i/kT \qquad (6)$$

Here $\nu_i$ is the wave number equals the number of vibration per sec divided by the velocity of light in cm per sec. It therefore has the physical dimensions of $cm^{-1}$. Ultimately, the total contribution to the vibrational entropy is the sum for all vibrations, taking into account any degeneracy where more than one mode of vibration has the same frequency.

This equation is derived as follows. According to Moore (1963), the vibrational energy $E$ is given as,

$$E = RT^2\partial \ln Q_{vib}/\partial T = \mathrm{L}h\nu/2 + \mathrm{L}h\nu \mathrm{e}^{-h\nu/kT}/(1 - \mathrm{e}^{-h\nu/kT})$$

Here $Lh\nu/2$ is the zero point vibrational energy $E_o$ remaining at absolute zero Kelvin, where L is Avogradro's number for the number of molecules in a mole. Thus, taking $h\nu/kT$ equal to x, as used above,

$$(E - E_o)/T = Rxe^{-x}/(1 - e^{-x})$$

$$(A_{vib} - E_o)/T = R\ln(1 - e^{-x}) \text{ since } A_{vib} = -kT\ln Q_{vib} = G_{vib}$$

Here $A$ and $G$ refer to Helmholtz and Gibbs energies.

So $S_{vib} = (E - G)/T = Rxe^{-x}/(1 - e^{-x}) - R\ln(1 - e^{-x})$ as given above.

Furthermore, the vibrational heat capacity is given as

$$\partial E/\partial T = C_{vib} = Rx^2e^{-x}/(1 - e^{-x})^2 = Rx^2/(e^x + e^{-x} - 2)$$
$$= Rx^2/2(\cosh x - 1), \text{ given } (e^x + e^{-x})/2 = \cosh x$$

Unlike the cases of translational and rotational entropy, it is not obvious that $S_{vib}T$ includes the vibrational kinetic energy. Indeed, it is only at temperatures above 470 K that $S_{vib}T$ exceeds $C_{vib}T$. Despite this, it is clear that each vibration of frequency $\nu_i$ contributes its own entropy.

$$S_v = \Sigma S\nu_i$$

In Figure 2, a flow diagram for computing entropy and Gibbs energy is given.

## Results

### *Data on molar entropies and absorption wavelengths of greenhouse gases*

We have established several concise relationships expressing entropy as a logarithmic function of an action ratio for translation and for rotation, with an exponential overlay on rotation by vibration.

$S_t = R\ln[e^{5/2}(@_t/\hbar)^3 z_t]$ (translation)

$S_r = R\ln[e(@_r/\hbar)^2/\sigma_r]$ (rotation – diatomic or linear molecule)

$S_r = R\ln[\pi^{1/2}e^{3/2}(@_A@_B@_C/\hbar^3)/\sigma_r]$ (rotation- polyatomic molecule)

$S_{vi} = Rx/(e^x - 1) - R\ln(1 - e^{-x})$, where $x = hc\nu_i/kT$ (for each vibrational mode)

For a diatomic molecules like nitrogen or oxygen making up most of the Earth's atmosphere we have for an action thermodynamics formulation,

$$S/N = s = k\{\ln[e^{7/2}(@_t/\hbar)^3(@_r/\hbar)^2 Q_e/(\sigma_r z_t)] + h\nu/kT/(e^{h\nu/kT} - 1) - \ln(1 - e^{-h\nu/kT})\} \quad (7)$$

or (given vibration is largely irrelevant for these molecules near 300 K, though not for carbon dioxide)

$$S/N = s = k\{\ln[e^{7/2}(@_t/\hbar)^3(@_r/\hbar)^2 Q_e/(\sigma_r z_t)] \quad (8)$$

We can compare this to the equivalent equation from statistical mechanics (Hill, 1960).

$$S/N = s = k\{\ln[(2\pi mkT/h^2)^{3/2}Ve^{5/2}/N] + \ln[(8\pi^2 kTI_r h^2)e/\sigma] + [(h\nu/kT)/(e^{h\nu/kT} - 1) - \ln(1 - e^{h\nu/kT})] + \ln\omega_e\} \quad (9)$$

It is clear that the translational action ratio $@_t/\hbar$ will vary as a function of temperature affecting velocity, but also with volume. In equations (7) and (8), action acts as a surrogate for the effects of both temperature and volume or density for translational entropy contained in equation (9). Normally, these variables are considered separately as shown in (9). At extremely low temperatures near absolute zero, the action ratio will tend to a minimum and the entropy will tend to zero, as required by the third law of thermodynamics. Near zero, only vibrational energy remains significant, expressed as the zero point vibrational energy of $h\nu/2$ per bond proposed as essential by Planck and Einstein.

The action-based forms of equations (7) and (8) are much more amenable for modeling and for understanding than the arbitrary forms of equation (9).

Given the results obtained here, there can be no doubt of the exact co-variation of entropy with action, accepting that the entropy contains additional terms related to kinetic energy or enthalpy as well as for symmetry. Causally, this must be based on the need for specific quantities of field energy to sustain action at a statistically stationary value, as appropriate for a particular kinetic environment and temperature. The statistical nature of entropy implicit in Boltzmann's and Gibbs' theories must also correspond with the relationship with action, discussed in Kennedy (2000). This statistical correspondence can be found in the space-filling dynamic nature of molecules subject to collisions, so that any complexion of low probability or high pressure involves chemical species occupying a comparatively small volume per molecule with low translational action as controlled by temperature; those complexions of higher probability or entropy occupy a comparatively large volume per molecule with higher translational action. This is also consistent with Shannon's information version of entropy, considering information as uncertainty and the capacity of a message to surprise (Shannon and Weaver, 1949). When an encounter is less frequent with greater diversity of species, surprise is more likely.

In Tables 2-5 calculations using these equations are indicated for individual contributions to the molar entropy of some common atmospheric and greenhouse gases. The computed entropy values at standard temperature and pressure compare very well with the rounded standard values in reference (Aylward and Findlay, 1974) obtained experimentally.

**Table 2a:** Entropy of linear gas molecules – translational, rotational

| Gas | MW | $I_t$ x$10^{40}$ g.cm$^2$ | $@_t/\hbar$ = $n_t$ | $S_t$=Rlne$^{5/2}$ x$(n_t)^3/z_t$] J K$^{-1}$ | Bond radius x$10^{10}$ cm | $I_r$ x$10^{40}$ g.cm$^2$ | $@_r/\hbar$ = $n_r$ | $S_r$= Rlne$(n_r)^2/\sigma$ J K$^{-1}$ | 1/λ cm$^{-1}$ | x = h ν/kT | $S_v$ | $Q_e$ | $S_e$ | ΣS |
|---|---|---|---|---|---|---|---|---|---|---|---|---|---|---|
| $H_2$ | 2.0000 | 98.83 | 104.7569 | 117.48 | 74 | 0.4580 | 1.8413 | 12.70 | - | - | - | 1 | 0 | 130.18 |
| $N_2$ | 28.0134 | 139.02 | 392.9013 | 150.45 | 110 | 14.235 | 10.2654 | 41.27 | - | - | - | 1 | 0 | 191.73 |
| $O_2$ | 31.9988 | 158.12 | 419.0198 | 152.06 | 121 | 19.590 | 12.0426 | 43.93 | 1580 | 7.63 | 0.04 | 3 | 9.13 | 205.16 |
| CO | 28.0110 | 138.33 | 391.9225 | 150.39 | 113 | 14.643 | 10.4115 | 47.27 | 2170 | 10.47 | 0.00 | 1 | 0 | 197.67 |
| NO | 30.0061 | 148.54 | 406.1341 | 151.28 | 115 | 16.555 | 11.0706 | 48.29 | 1904 | 9.188 | 0.01 | 4 | 11.5 | 211.12 |
| $CO_2$ | 44.0099 | 217.42 | 491.3535 | 156.03 | 244 | 79.665 | 24.2846 | 54.72 | See | below | 2.99 | 1 | 0 | 214.61 |
| $N_2O$ | 44.0134 | 215.90 | 489.6496 | 155.94 | | 66.9 | 22.2550 | 59.90 | See | below | 3.05 | | 0 | 218.89 |

Data obtained from Herzberg (1945, 1950) and Aylward and Findlay (1974)

**Table 2b:** Triatomic gases translational and rotational entropy

| Gas | MW | $I_t$ x$10^{-40}$ g.cm$^2$ | $@_t/\hbar$ | $S_t$ JK$^{-1}$ | $I_{rA}$ x$10^{-40}$ | $I_{rB}$ g.cm$^2$ | $I_{rC}$ x$10^{40}$ | $@_{rA}/\hbar$ | $@_{rB}/\hbar$ | $@_{rC}/\hbar$ | $\sigma_r$ | $S_r$ JK$^{-1}$ | Point group |
|---|---|---|---|---|---|---|---|---|---|---|---|---|---|
| $H_2O$ | 18.0154 | 88.372 | 313.2678 | 144.80 | 1.024 | 1.920 | 2.947 | 2.7533 | 3.7699 | 4.6709 | 2 | 43.74 | $C_{2v}$ |
| $H_2S$ | 34.080 | 167.18 | 430.8673 | 152.75 | 2.667 | 3.076 | 5.845 | 4.4435 | 4.7721 | 6.5779 | 2 | 52.52 | $C_{2v}$ |
| $O_3$ | 47.9982 | 235.45 | 511.3358 | 157.02 | 7.877 | 62.865 | 70.900 | 7.6366 | 21.5796 | 22.9104 | 2 | 79.94 | $C_{2v}$ |
| $SO_2$ | 64.0628 | 314.25 | 590.7399 | 160.62 | 13.807 | 81.328 | 95.356 | 10.1103 | 24.5377 | 26.5697 | 2 | 84.58 | $C_{2v}$ |

**Table 2c:** Triatomic gases vibrational entropy

| $H_2O$ | Wave number | x= $hc\nu_i/kT$ | $S_{vi}$ | | $CO_2$ | Wave number | x= $hc\nu_i/kT$ | $S_{vi}$ | Degen | ∑Svi |
|---|---|---|---|---|---|---|---|---|---|---|
| $A_1$ | 3652 | 17.6235 | <0.0001 | | $\sigma_g^+$ | 1388 | 6.6981 | 0.0790 | 1 | 0.0790 |
| $A_1$ | 1595 | 7.6970 | 0.0329 | | Π | 667 | 3.2188 | 1.4547 | 2 | 2.9093 |
| $B_2$ | 3756 | 18.1254 | <0.0001 | | $\sigma_u^+$ | 2349 | 11.3356 | 0.0012 | 1 | 0.0012 |
| Total | | Total | 0.033 | | | | | | Total | 2.9895 |
| | | | | | | | | | | |
| $H_2S$ | Wave number | x= $hc\nu_i/kT$ | $S_v$ | | $N_2O$ | Wave number | x= $hc\nu_i/kT$ | $S_v$ | | |
| $A_1$ | 2615 | 12.6193 | 0.004 | | ∑ | 2224 | 10.7324 | 0.0002 | | |
| $A_1$ | 1183 | 5.7088 | 0.1856 | | ∑ | 1285 | 6.2011 | 0.1216 | | |
| $B_2$ | 2626 | 12.6723 | <0.0001 | | Π | 589 | 2.8423 | 1.4627 | | |
| | | Total | 0.1860 | | Π | 589 | 2.8423 | 1.4627 | | |
| | | | | | | | Total | 3.0473 | | |
| $O_3$ | | x= $hc\nu_i/kT$ | | | $SO_2$ | | x= $hc\nu_i/kT$ | | | |
| $A_1$ | 1110 | 5.3565 | 0.2504 | | $A_1$ | 1151 | 5.5544 | 0.2117 | | |
| $A_1$ | 705 | 3.4021 | 1.2561 | | $A_1$ | 518 | 2.4997 | 2.5715 | | |
| $B_2$ | 1042 | 5.0284 | 0.3301 | | $B_2$ | 1352 | 6.5244 | 0.0919 | | |
| | | | 1.8367 | | | | | 2.8751 | | |

**Table 3a:** Polyatomic molecules

| Gas | MW | $I_t$x$10^{40}$ g.cm$^2$ | $@_t/\hbar$ | $S_t$ J K$^{-1}$ | $I_{rA}$ x$10^{40}$ | $I_{rB}$ g.cm$^2$ | $I_{rC}$ x$10^{40}$ | $@_{rA}/\hbar$ | $@_{rB}/\hbar$ | $@_{rC}/\hbar$ | σ | $S_r$ J K$^{-1}$ | Point group |
|---|---|---|---|---|---|---|---|---|---|---|---|---|---|
| $NH_3$ | 17.031 | 83.543 | 27.9881 | 144.10 | 2.9638 | 2.9638 | 4.5176 | 4.6841 | 4.6841 | 5.7830 | 3 | 48.36 | $C_{3v}$ |
| | | | | | | | | | | | | | |
| Species | Wave number | x= $hc\nu_i/kT$ | $S_v$ | | | | | | | | | | |
| $A_1$ | 3337 | 16.1034 | 0.0001 | | | | | | | | | | |
| $A_1$ | 950 | 4.5844 | 0.4785 | | | | | | | | | | |
| E | 3447 | 16.6343 | 0.0001 | | | | | | | | | | |
| E | 1627 | 7.8514 | 0.0287 | | | | | | | | | | |
| | | | 0.5074 | | | | | | | | | | |

**Table 3b:** Polyatomics

| **Gas** | **MW Daltons** | **$I_t$ x10$^{40}$ g.cm$^2$** | **$@_t/\hbar$** | **$S_t$ JK$^{-1}$** | **$I_{rA}$** x10$^{40}$ | **$I_{rB}$ g.cm$^2$** | **$I_{rC}$** x10$^{40}$ | **$@_{rA}/\hbar$** | **$@_{rB}/\hbar$** | **$@_{rC}/\hbar$** | **σ** | **$S_r$ JK$^{-1}$** | **Point group** |
|---|---|---|---|---|---|---|---|---|---|---|---|---|---|
| $CH_4$ | 16.401 | 78.678 | 295.6225 | 143.35 | 5.27 | 5.27 | 5.27 | 6.2461 | 6.2461 | 6.2461 | 12 | 42.263 | $T_d$ |
| $CFCl_3$ | 137.37 | 673.84 | 865.0408 | 170.13 | 340.35 | 340.35 | 799.71 | 50.1970 | 50.1970 | 62.433 | 3 | 107.59 | $C_{3v}$ |
| $CF_2Cl_2$ | 120.91 | 593.12 | 811.5780 | 168.54 | 203.73 | 318.00 | 375.66 | 38.8364 | 48.5206 | 52.737 | 2 | 107.13 | $C_{2v}$ |
| $CF_3Cl$ | 104.46 | 512.41 | 754.3356 | 166.71 | 146.32 | 251.58 | 251.58 | 32.9126 | 43.1566 | 43.1566 | 3 | 99.745 | $C_{\infty v}$ |
| $CH_4$ | Wave number | x= $hc\nu_i/kT$ | $S_v$ | Degen. | $\Sigma S_v$ | | $CFCl_3$ | Wave Number | x= $hc\nu_i/kT$ | $S_v$ | Degen. | $\Sigma S_v$ | |
| A | 2914 | 14.063 | 0.0001 | 1 | 0.0001 | | A | 1085 | 5.2359 | 0.2773 | 1 | 0.2733 | |
| E | 1526 | 7.3640 | 0.0441 | 2 | 0.0882 | | A | 535 | 2.5818 | 2.4105 | 1 | 2.4105 | |
| T | 3020 | 14.575 | 0.0001 | 3 | 0.0002 | | A | 350 | 1.6890 | 4.8792 | 1 | 4.8792 | |
| T | 1306 | 6.3034 | 0.1113 | 3 | 0.3340 | | E | 847 | 4.0874 | 0.7208 | 2 | 1.4416 | |
| | | | | Total | 0.4225 | | E | 394 | 1.9013 | 4.1210 | 2 | 8.2420 | |
| | | | | | | | E | 241 | 1.1630 | 7.5121 | 2 | 15.024 | |
| | | | | | | | | | | | Total | 32.275 | |
| $CF_2Cl_2$ | cm$^{-1}$ | x= $hc\nu_i/kT$ | $S_v$ | $CF_3Cl$ | Band | cm$^{-1}$ | x= $hc\nu_i/kT$ | $S_v$ | Degen. | $\Sigma S_v$ | | | |
| A | 1101 | 5.3131 | 0.2598 | | A | 1105 | 5.3324 | 0.2556 | 1 | 0.2556 | | | |
| A | 667 | 3.2188 | 1.4547 | | A | 781 | 3.7689 | 0.9344 | 1 | 0.9344 | | | |
| A | 458 | 2.2108 | 3.2297 | | A | 476 | 2.2970 | 3.0163 | 1 | 3.0163 | | | |
| A | 262 | 1.2643 | 6.8968 | | E | 1212 | 5.8488 | 0.1646 | 2 | 0.3293 | | | |
| A | 322 | 1.5539 | 5.4387 | | E | 563 | 2.7169 | 2.1667 | 2 | 4.3334 | | | |
| B | 902 | 4.3528 | 0.5796 | | E | 350 | 1.6890 | 4.8792 | 2 | 9.7584 | | | |
| B | 437 | 2.1088 | 3.4981 | | | | | | Total S | 18.627 | | | |
| B | 1159 | 5.5930 | 0.2048 | | | | | | | | | | |
| B | 446 | 2.1523 | 3.3804 | | | | | | | | | | |
| | | Total S | 24.945 | | | | | | | | | | |

An interesting feature of the data calculated for Tables 4 and 5, but rarely nechanistically considered, is the significantly greater entropy related to translational action compared to rotational and vibrational action for all gas molecules at 298 K. The lower the pressure of a particular gas the greater is this discrepancy. Indeed, most of the heat required to raise the temperature of the gas from zero to 298 Kelvin under standard conditions is devoted to sustaining translational action, with only a small proportion of molecules exhibiting any vibrational action and entropy at all. If we consider that the relative action states $@/\hbar$ for carbon dioxide at standard temperature and pressure calculated here are 491 for translation, 24 for rotation and a much lower number for excited vibration, indicating a very low proportion of molecules excited with infra-red quanta, it is reasonable to conclude that the size of the quanta associated with rotation and translational action states have correspondingly lower frequencies. This would appear to place them in the microwave and the radio-wave range of the electromagnetic spectrum – relatively cold or dark energy.

Thus, although heat is required to generate translational entropy, could its actual form be considered as gravitational work, although for sub-orbital molecules with quanta of relatively high frequency. If so, we could consider translational action and entropy as indicating the quantity of heat required to reach $T$ degrees Kelvin in the gravitational field, including that normally regarded as pressure-volume work.

**Table 4:** Summary of total entropy terms

| **Gas** | **$S_t$ $JK^{-1}$** | **$S_r$ $JK^{-1}$** | **$S_v$ $JK^{-1}$** | **ΣS $JK^{-1}$** | **(A&F) $JK^{-1}$** | **IR spectrum Wavelength μm (3n-6)** |
|---|---|---|---|---|---|---|
| $H_2O$ | 144.80 | 43.74 | 0.033 | 188.6 | 189 | 2.662, 2.738, 6.270 |
| $CO_2$ | 155.94 | 54.715 | 2.99 | 213.6 | 214 | 4.257, 7.2046, 14.993 |
| $H_2S$ | 152.81 | 52.54 | 0.19 | 205.5 | 206 | 3.808, 3.824, 8.453 |
| $N_2O$ | 155.94 | 59.80 | 3.05 | 218.9 | 220 | 4.446, 7.782, 16.978 |
| $O_3$ | 157.02 | 79.94 | 1.84 | 238.8 | 239 | 9.009, 9.597, 14.184 |
| $SO_2$ | 160.62 | 84.58 | 2.875 | 248.1 | 248 | 7.396, 8.688, 19.305 |
| $NH_3$ | 144.10 | 48.36 | 0.507 | 193.0 | 192 | 2.901, 2.997, 6.146, 10.526 |
| $CH_4$ | 143.36 | 42.26 | 0.423 | 186.0 | 186 | 3.311 (2), 3.432 (3), 6.553 , 7.657 (3) |
| $CFCl_3$ | 170.14 | 107.59 | 32.275 | 310.0 | 310 | 9.217, 11.806(2), 18.692, 25.381(2), 28.571, 41.494(2) |
| $CF_2Cl_2$ | 168.545 | 107.137 | 24.945 | 300.6 | 301 | 8.628, 9.083, 11.086, 14.999, 21.834, 22.421, 22.883, 31.056, 38.168 |
| $CF_3Cl$ | 166.721 | 99.749 | 18.627 | 285.1 | 286 | 8.251(2), 9.050, 12.804, 17.762(2), 21.008, 28.571(2) |
| $O_2$ | 162.07 | 43.93 | 0.035 | 206.0 | 205 | 6.329 |
| CO | 150.31 | 47.19 | 0.0025 | 197.5 | 198 | 4.608 |
| NO | 162.69 | 48.39 | 0.0087 | 211.6 | 211 | 5.252 |
| $H_2$ | 117.48 | 12.70 | - | 130.2 | 131 | - |
| $N_2$ | 150.45 | 41.27 | - | 191.7 | 192 | - |
| A | 154.84 | - | - | 154.8 | - | |

We must regard the heat that was required to melt and vaporise the carbon dioxide molecules as having performed configurational work, either on separating the molecules or pressure-volume work of lifting the atmosphere. This work-heat consumption identified by Clausius (1875) explains the fact that not all the heating included in the entropy function contributes to sensible heat just raising the temperature as increased kinetic energy, consistent with the heat capacity of each molecule. Similar conclusions can be drawn for the melting of ice to water and its subsequent vaporization. Such interpretations regarding inter-conversions of heat and work require further investigation and this may be facilitated using the quantum features of the action approach.

***Considering water's phase changes***

Of all the atmospheric gases considered here, only water exists in the atmosphere on Earth as gas, liquid and solid. This erratic cycle is very apt for illustrating significant changes in entropy states associated with changes in phase. Most of the permanent gases in the atmosphere only exist as vapors. As a result, their changes in entropy refer only to changes in kinetic and potential energy corresponding to changes in enthalpy and in free energy as a response to changes in temperature and pressure respectively. However, water has highly significant changes in action and entropy in the atmospheric weather cycle, with corresponding consumption or release of heat.

The total thermal capacity to bring a mole of water to vapor at 298.15 K and 1 atmosphere pressure (were this possible) is 56.2 kJ. Of this, only 7.4 kJ can be attributed to its heat capacity as increased kinetic energy per mole over the temperature range, with 5.8 kJ required for melting and 44.0 kJ to vaporisation at 298 K. Almost 90% of the solar heat absorbed by water as vapor in the atmosphere is available for release in the two phase transitions of forming snow or hail. Although these facts are well known and the major warming possible during atmospheric condensation of water vapor is understood, this could also be a fruitful area for further investigation using the action-entropy theory.

It is of interest from Table 4 that water has only minor vibrational entropy – lower even than that of oxygen, mainly at a very similar short infrared wavelength (6.270 and 6.329 μm) – potentially allowing overlap for emission or absorption. On this basis, there might seem to be only a weak case to consider water a major greenhouse gas, although this is customary. In fact, the converse is true as its low actual excitation of vibration, particularly by the two shorter wavelengths for water is that there remains a high population of water molecules able to be excited by radiation from the Earth's surface. However, the reversible latent heat of vaporisation of water released when it condenses in clouds or at dewpoint is also an important factor for heat transfer in the atmosphere.

It is also apparent that the actual vibrational entropy of methane is only about one-sixth that exhibited by carbon dioxide and nitrous oxide. Presumably the low vibrational entropy can also be related to a higher residual absorptivity of methane, but it is only slightly statistically enhanced by being poorly excited at this temperature. Only a very low proportion of methane as well as water molecules are excited by infrared radiation at equilibrium under ambient temperature conditions, as shown by their low vibrational entropies at 298.15 K.

Shown in Tables 3a,3b and 4, the organo-halogens such as Freon 11 ($CFCl_3$) that have been withdrawn from use under the Montreal Protocol have an exceptionally large vibrational contribution to entropy. Replacing the hydrogen atoms of methane with these two halogen atoms also significantly increases both the rotational and vibrational entropy; this should lessen absorptivity in the longer infrared region significantly since this is relatively excited at 298 K as a result of longer bond lengths and greater ease of dissociation of atoms. According to Glasstone (1951), the formula for calculating vibrational entropy strictly applies only to divalent molecules. However, this cautionary note may not be required. By applying the formula to each bond separately and summating as shown in Table 3b, including any degeneracy, the agreement with experimentally determined entropies using the Third law approach is just as good as for other molecules where only translational and rotational contributions are significant.

Nitric oxide (NO), although not a greenhouse gas with only one vibrational line in the short-wave infrared (5.252μm), is included for comparison, as are CO (4.608 μm) and $O_2$ (6.329 μm) (Tables 2, 4). For nitric oxide (NO) a large discrepancy in total entropy between the data calculated here from translation, rotation and vibration would occur if the electronic ($Q_e = 4$) term was neglected, a result of its free radical nature containing unpaired electrons; these add $R\ln 2^2$ or 11.53 extra entropy units per mole, giving a total value of 211.1, in agreement with the Aylward and Findlay value (1974).

The Sackur-Tetrode equation includes a term for the electronic partition function ($Q_e$) . In the case of $O_2$, the ground state electronic partition function $Q_e$ is 3 at STP because this molecule has two unpaired electrons that can have their two spins oriented three ways with respect to the nuclear spin – both up, both down and oppositely. Because they can be distinguished, the three different oxygen species have three times the volume per particle, affecting their action because of the greater radial separation than if only a single species existed. This gives an additional electronic entropy contribution of $S_e = R\ln Q_e$ or $R\ln 3$.

However, the fact that a mole of oxygen will contain one-third of a mole of each species must be also be considered in estimating the total entropy. This $Q_e$ factor is included in the Sackur-Tetrode equation as $S_t = R\ln[Q_e e^{5/2}(V/N)(2\pi mkT)^{3/2}/(h^3)]$, or in the action form of the equation as $R\ln[e^{5/2}(@_t/\hbar)^3 Q_e/z_t]$, which is equal to $R\ln Q_e - R\ln z_t + 2.5R + 3R\ln(@_t/\hbar)$. Thus asymmetry ($Q_e$) increases entropy by increasing spatial distance between molecular interactions and symmetry ($z_t$) decreases it by reducing spatial distances– less field energy is needed to sustain a symmetrical molecule than an asymmetrical one.

Because of the statistical variation in quanta for rotational and translational fine structure, the actual vibrational spectra are not confined to these spectral lines but distributed around these wavelengths. The spectra may be sharpened by cooling the gas and this is usually done when testing the theory with data.

The data in Tables 2-4 are calculated for standard conditions of temperature and pressure. To adjust these results to the actual gas pressures in the atmosphere at the same temperature, only the translational action and entropy will vary. In terms of centimeter-gram-second (cgs) units, the pressure, is equal to $kTa^{-3}$ or $kT/8r^3$ at $1.013\text{x}10^5$ pascals, being the product of the mass of air per square cm of the earth's surface (ca. 1 kg) and the acceleration of gravity (9.807 $\text{m.sec}^{-2}$).

It should be understood that entropy is a dimensionless number corresponding to its role in probability, given that it expresses a total thermal capacity per degree of temperature – a ratio of extensive and intensive measures of energy. It is also instructive to be aware that the product of entropy and absolute temperature (*ST*) is always a significant multiple of the kinetic energy since that is merely one of its components – to its kinetic energy must be added the sustaining field energy corresponding to decreases in free energy from absolute zero while heating the molecules, absorbing any heat that becomes latent during this process and in doing any work such as breaking H-bonded aggregated structures or pressure-volume work against the atmosphere. In this connection, much of the magnitude of *ST* is generated together with increased enthalpy during phase changes when parent solid or liquid matter is melting or vaporizing. Gibbs energy does not change when these reversible processes occur isothermally. The chemical potential of the liquid water is equilibrated with that of the vapour at the boiling temperature, the increase in the enthalpy on vaporization being effectively an increase in internal entropies associated with increasing the internal vibration and rotation of the de-clustered water molecules. Such increases in internal action and entropy are actually increases in enthalpy.

Any decrease in the density of a chemical substance such as the expansion of a solid, liquid or gas will also increase *ST* as its Gibbs free energy decreases. For example, liquid water gradually changes its state during heating from large H-bonded clusters of about 30 water molecules just above freezing to fewer than half that number per cluster just below boiling temperature (Kennedy, 2001), above which the clusters completely dissociate. The variable action of these flickering clusters between zero Celsius and 100 C could be calculated and the changes in entropy estimated.

Some results calculated for the actual sea level pressures (ppmv) of all atmospheric gases at the standard temperature of 298.15 K are given in Table 5. Entropy values are given per mole of each substance, using Boltzmann's constant *k* multiplied by Avogadro's number N as the unit value and the product [entropy x temperature, *ST*] estimated for each as a proportion of the total.

The total entropic energy as Σ*ST* in air at sea level is about 2.4 MJ per cubic metre. It is obvious that the very dilute gases like nitrous oxide and methane have a relatively large translational entropy compared to the major gases and therefore need more heat per molecule to bring them to this temperature and pressure. The majority of the heat required (*ST*) to raise the atmosphere is absorbed into the fields of only three different molecules – nitrogen, oxygen and water. Given the reversible phase changes available to water, most of its maximum entropic energy is made available during condensation as part of the hydrological cycle. Roderick *et al*. (2013) have estimated that for a warming of 2.8 K, the atmospheric content of water would increase from an equivalent liquid column of 30 mm to 35.9 mm, or 7% per degree K of warming. According to Table 5, by proportion alone, this would amount to 4,241 J per cubic metre of air at the surface of extra heat required. However, an exact calculation would need to consider the diminution of the translational entropy per molecule as a result of its increased concentration.

**Table 5:** Summary of total entropy and entropy-temperature terms in the real atmosphere at 298.15 K

| Gas | Pressure (atm) | $S_t$ | $S_r$ | $S_v$ | ΣS Total | ΣS/ mole STP | J/Mole of air/K | J per m$^3$ |
|---|---|---|---|---|---|---|---|---|
| $H_2O$ | 0.00775 | 185.29 | 43.74 | 0.033 | 229.1 | 188.6 | 529.37278 | 21,637.477187 |
| $CO_2$ | 0.000397 | 215.51 | 54.72 | 2.99 | 273.2 | 213.6 | 32.337468 | 1,321.755203 |
| $H_2S$ | 0.0000000002 | 338.49 | 52.52 | 0.19 | 391.2 | 205.5 | 0.0000023 | 0.000003 |
| $N_2O$ | 0.000000325 | 280.24 | 59.90 | 3.05 | 342.2 | 218.9 | 0.0331588 | 1.355326 |
| $O_3$ | 0.0000000266 | 302.13 | 79.94 | 1.84 | 383.9 | 238.8 | 0.0032865 | 0.134332 |
| $SO_2$ | 3 x 10$^{-10}$ | 343.01 | 84.58 | 2.875 | 430.5 | 248.1 | 0.0000039 | 0.000159 |
| $NH_3$ | 5x10$^{-10}$ | 322.23 | 48.36 | 0.507 | 371.1 | 193.0 | 0.0000553 | 0.002260 |
| $CH_4$ | 0.0000017 | 272.21 | 42.26 | 0.423 | 314.9 | 186.0 | 0.1596086 | 6.523810 |
| $CFCl_3$ | 0.00000000026 | 350.00 | 107.59 | 32.28 | 489.9 | 310.0 | 0.0000380 | 0.000006 |
| $CF_2Cl_2$ | 0.00000000055 | 345.90 | 107.14 | 24.945 | 478.0 | 300.6 | 0.0000784 | 0.000001 |
| $CF_3Cl$ | 0.0000000001 | 353.08 | 99.75 | 18.627 | 471.5 | 285.1 | 0.0000141 | 0.000002 |
| $O_2$ | 0.2095 | 165.05 | 43.93 | 0.035 | 209.0 | 205.1 | 13,054.6468 | 533,593.023660 |
| CO | 0.00000015 | 279.58 | 47.19 | 0.0025 | 326.8 | 197.5 | 0.0146153 | 0.597382 |
| NO | 3x10$^{-10}$ | 333.56 | 48.39 | 0.0087 | 382.0 | 199.6 | 0.0000034 | 0.000139 |
| $H_2$ | 0.0000005 | 238.11 | 12.70 | - | 250.8 | 130.2 | 0.0373880 | 1.528190 |
| $N_2$ | 0.78084 | 152.45 | 41.27 | - | 193.7 | 191.7 | 45,094.8022 | 1,843,195.930604 |
| A | 0.00934 | 193.70 | - | - | 193.7 | 154.8 | 539.400466 | 0.0442339 |
| | | | | | | | **Total** | **2,399,758.372498** |

In Table 5, this operation is illustrated for the densest surface layer of the atmosphere only. If the temperature and pressure of the gas is known, a similar calculation can easily be repeated at all altitudes in the troposphere or the stratosphere and the results integrated to give the total heat capacity of the atmosphere, a task that others are invited to perform. For such calculations it is convenient to use suitable computer programs. Fully annotated outlines of these programs are available on request to the corresponding author.

## Discussion

### *Phase space as action space*

As shown in the equations above, the mean molecular Gibbs energy can always be expressed as a variable of the relative action ratio $@/\hbar$ alone. Entropic energy ($sT$) also includes kinetic energy and the capability for pressure-volume work that are considerd as enthalpy. This extends to vibrational and electronic states although their contribution to atmospheric gases near the surface of the Earth at ambient temperatures is usually relatively small, as shown in the tables. Using the action model, the negative relationship between free energy and the entropy is more clearly revealed. Paradoxically, "free energy" is not real energy, but actually denotes its absence; it is better viewed as a system's action potential showing its capacity to accept thermal energy by increasing its action and sustaining field energy, building more complex and diverse structures whilst doing external work such as expanding the atmosphere or lifting weights in a gravitational field. To the extent that cooling gravitational work is done that increases the free energy, heat may re-emerge later if reverse work is done on the molecules of the system. This reversibility is the essence of the second law of thermodynamics in action theory. We also consider that referring to this process of increasing entropy as one of increasing *disorder* is misleading (Kennedy, 2001). Diversity would be more appropriate term.

Given that the relative action or mean quantum number $@/\hbar$ can be expressed simply as a function of the particle's mass, its radial separation and the square root of the temperature affecting velocity, all of the paradoxes regarding entropy such as its lack of change during the mixing of equal volumes of identical gases *versus* the change when two distinguishable gases are mixed at the same final pressure and temperature are easily resolved. Each of the distinguishable molecules now occupy twice the space as formerly, increasing their action accordingly, whereas identical gases must remain in the same space with the same action as before mixing.

In principle, the suggestion to calculate entropy from the logarithm of the translational and rotational action (including its modification by vibration) is not new. Gibbs identified the significance of action in his classical text (1902), describing it as the *extension in phase* ($V_{pq}$), claiming that "the quantity …. which corresponds to entropy is log *V*, the quantity *V* (not volume) being defined as the extension in phase". So we can conclude that, according to Gibbs even before Planck identified his quantum of action, any equi-potential contour in phase space of equal translational action ($V_p$ x $V_q$ = *mv* x *r*) would also correspond to states of equal translational entropy. In effect, changes in the momentum *mv* and a linear coordinate *r* would lead to no change in their product action and its logarithm, entropy. We can now recognise such contours as adiabatics differing by a minimum of Planck's quantum of action *h*, giving a scale for estimating maximum uncertainty in momentum or position.

No claim is made here that the action model is inherently more accurate than the classical methods of statistical mechanics or that the formulae given here for ideal gases apply without corrections under all conditions of temperature or pressure. But the results are easily obtained from primary data and are surprisingly accurate, even for vibrational entropy. This suggests that the action method will have strong heuristic value – not only for climate science but also for theoretical and experimental purposes in all branches of chemistry and physics. For example, this action revision of the nature of entropy and free energy and the interaction between internal and translational action states has the potential to advance reaction rate theory and many other processes occurring in the liquid state, including those of life systems. In this area, the translational action will play a special role, since it is closely related to changes in Gibbs energy for molecular trajectories from the chemical potential of free reactants through reversibly activated transition states to the chemical potential of products (see Kennedy, 2001, chapters 4, 6).

***Boltzmann's realistic collision model of entropy***

Strongly relevant to the equations for entropy based on logarithmic functions of action given in this paper is the approach used by Ludwig Boltzmann; he derived an equation for entropy using his *H*-theorem by considering the mathematical behavior of a collision integral using a realistic model. This theorem was based on integrating the average effect on a single molecule of collisions with all the other molecules of the gas, spontaneously increasing its entropy whenever commencing with a more ordered state. This led him to essentially the same equations for entropy as those of Gibbs, while claiming "that the mechanical basis is necessary to illustrate the abstract equations", despite the current of opinion at the time from Mach and others directed against the existence of molecules. For example, Boltzmann (1896) gives an expression for the integral of the sum of the entropies of the masses in the volume elements as [$R\ln(\rho^{-1}T^{3/2})$ + const.], where ρ is the number density of gas molecules. We can observe that this result only lacks a suitable divisor in the logarithmic term required to remove the physical dimensions of action per unit mass.

In his collision theory, Boltzmann establishes that the quantity *H*, which can be identified with the negative of the entropy, must always decrease with time or remain constant – if one assumes that the velocity distributions of colliding molecules are uncorrelated. *H* remains constant (i.e. $dH/dt$ is zero) only when the gas attains a most probable velocity distribution, corresponding to that derived by Maxwell, a process requiring collisions. This approach related entropy with probability and Planck's derivation of the quantum of action for radiation (1913) acknowledges Boltzmann's statistical theory of isothermal entropy ($S = k\ln W$). Relating this to the formula for entropy given in this paper, we conclude that the position of equilibrium where *H* is constant, must correspond to an isothermal state where the action is also stationary, although they will both be maximal.

In action theory (Kennedy, 2001) we have extended this realistic model (Fig. 1), with molecular symmetry reducing the need for field energy because of the shorter free path for energy between identical molecular structures. Conversely, asymmetry maximizes the free path length between molecules of field energy and its magnitude in sustaining a system's structures. The heat capacity of such diverse systems and their ability or need to store field energy at a given temperature is therefore greater. For all practical purposes, the only variable affecting the entropy is the action; more diverse systems will automatically display greater relative action because of the greater separation of identical molecular structures, but these will also have a richer energy content for a given temperature.

In general, only differences in action and entropy between alternate states are of thermodynamic interest since the absolute entropy is not required. We have shown that for translation the entropy change per mole given a change of state (1=>2) can be expressed as a function,

$$\Delta S_{t2\text{-}t1} = R\ln[(e^{5/2}@_{t2}/\hbar)^3/z_t] - R\ln[e^{5/2}(@_{t1}/\hbar)^3/z_t] = 3R\ln[(@_{t2}/(@_{t1})] \quad (10)$$

assuming that the translational symmetry factor $z_t$ remains the same between state 1 and state 2. Note that the kinetic or enthalpic aspect of entropy drops out. So equation (10) for the change in entropy gives statistical effects only. Therefore, the precise choices of the symmetry factor or of the translational radius are only of significance for estimating the absolute entropy. Because of the statistical nature of momentum and position in phase or action space, assigning an exact value to the most probable radius or symmetry factor to each molecule is impossible since they will fluctuate around statistical mean values. However, the amplitude of the fluctuations is of interest, since these will control rates of transition.

***Under isothermal conditions Gibbs energy varies with translational action and entropy***

Given that the third law of thermodynamics states that the entropy at the temperature of absolute zero is zero, this would require that the action ratio $@_t/\hbar$ for a gas at this minimum temperature must be slightly less than one, if it could exist as such, since the translational entropy can then be considered as equal to $5/2R + R\ln[(@_t/\hbar)^3/z_t]$, equivalent to the Gibbs expression for entropy of $ST = H - G$ where $G$ is the free energy or work potential of a monatomic gas at constant pressure. This suggests that the magnitude of the function $RT\ln[(@_t/\hbar)^3/z_t]$ has the same value as the free energy, although opposite in sign, so that $G = -RT\ln[(@_t/\hbar)^3/z_t]$ or $RT\ln[z_t(\hbar/@_t)^3]$ and $\Delta G = 3RT\ln[(@_{tr}/(@_{tp})]$ for changes in action state at constant temperature.

Incidentally, the total entropy change during an isothermal chemical reaction at $T$ includes contributions from changes in translational action and entropy – $\delta(S_t T)$ is equal to the change in Gibbs chemical potential – as well as changes in the internal action and entropy representing changes in enthalpy as a result of revised bonding energies. It is important to understand that the enthalpy term designated $H$ refers to the sensible heat that tends to change temperature. Thus, if a chemical reaction results in products where atoms or electrons are more firmly bound with shorter radii, the reduced potential energy will be compensated by increased internal kinetic energy and equal quantities of emitted quanta, resulting in a release of heat as a reduction in Gibbs energy and an increase in entropy and decrease in Gibbs energy of the surrounding system. This change in heat content as a function of changes in chemical bonding will be reflected in the variation in zero-point energies of all vibrations possible. In the absence of such chemical reactions, the enthalpy change can be measured by the changes in kinetic energy and pressure-volume work alone. This is obviously true with monatomic noble gases like argon that do not undergo chemical reaction.

It is important to note that, within limits, the internal entropy for rotational and vibrational states is a function of temperature only. It is unaffected by changes in concentration, except at very high densities, in contrast to translational states. A low concentration or pressure corresponds to a high action state of greater entropy. So at constant temperature, changes in free energy are purely a function of changes in translational action states, since internal entropy or enthalpy remains constant, or fluctuate around a stable mean value though variations in internal states by absorption or emission of radiation and re-equilibration with translational states. This is entirely consistent with chemical work processes being directly mediated by translational inertia and pressure, such as pressure-volume work.

Consequently, the free energy or action potential can be considered simply as an inverse logarithmic function of the translational action and entropy at a given temperature $T$. The lower the initial action, the greater the Gibbs energy and action potential, indicating a spontaneous propensity for thermal energy to be absorbed as latent heat and the action to

increase – thus providing the dynamic functional basis for the second law of thermodynamics. As a measure of the energy capacity of molecular systems, the translational entropy must also include terms for kinetic energy and pressure-volume work, indicated by the 2.5$R$ or exponential ($e^{5/2}$) term. But these, together with the translational symmetry factor, cancel out for changes at constant temperature.

We can then write that

$$G = H - ST \quad = \quad H - (S_t T + S_r T + S_{vi} T)$$

The enthalpy (H) is a term always referring to the sensible heat in a system that can be sensed with a thermometer and related to the kinetic energy of its molecules. The entropic energy $ST$ differs in that it is only partly indicative of sensible heat, but includes the potential energy stored in work such as thermodynamic work in molecular systems, or gravitational work. According to the Carnot principle, in a reversible system this work can reappear as sensible heat, raising the temperature. This potential source of extra warming certainly applies to gases in the Earth's atmosphere. Indeed, it is responsible for much of the heat transfer to higher latitudes, released by frictional processes on the Earth's surface.

For monatomic gases we can rewrite this classic equation taught to all students using the algorithms developed here as follows.

$$G = H - ST$$

$$-RT\ln[(@_t/\hbar)^3/z_t] = 1.5RT + RT - RT\ln[e^{5/2}(@_t/\hbar)^3/z_t]$$

For the main diatomic gases in the atmosphere nitrogen and oxygen at ambient temperatures by including rotational entropy but where we can neglect vibrational entropy, we will have the following.

$$-RT\ln\{[(@_t/\hbar)^3 Q_e/z_t][(@_r/\hbar)^2/\sigma_r]\} = 2.5RT + RT - \mathrm{RT}\ln\{[e^{7/2}(@_t/\hbar)^3 Q_e/z_t]\,[(@_r/\hbar)^2/\sigma_r]\}$$

Disallowing a role for the enthalpy of chemical reactions at ambient temperatures in the troposphere we have,

$$G = H - ST = 3.5RT - ST$$

Alternatively, we can write for Helmholtz energy in constant volume conditions a modified equation, varying only slightly in the *RT* or *PV* term.

$$-RT\ln\{[e(@_t/\hbar)^3 Q_e/z_t][(@_r/\hbar)^2/\sigma_r]\} = 2.5RT - \mathrm{RT}\ln\{[e^{5/2}(@_t/\hbar)^3 Q_e/z_t]\,[(@_r/\hbar)^2/\sigma_r]\}$$

$$A = E - ST = 2.5RT - ST = G - RT$$

For polyatomic molecules or at temperatures where vibrational entropy and energy are more relevant, it is simple to add the vibrational entropy terms to both sides of the equation.

Physicists and chemists have preferred to consolidate the different kinds of entropy into the single term *S*, perhaps for simplicity. They have then relied on differential calculus to establish thermodynamic relationships. However, this has partly concealed its true nature, obscuring the relationship between the partitioned kinds of action. Furthermore, action thermodynamics proposes that the external action (translation) and the internal actions (rotation, vibration, electronic orbital) defining the respective translational and internal free energies or entropies can sometimes be considered as physically opposed [5] during collisions and able to vary independently. Since these can now be calculated more easily and integrated numerically if required, the need for solutions based on calculus or differential equations is diminished.

In his engaging book on statistical mechanics, Schrödinger (1946, p. 52) derived the relationship for the Gibbs potential as $nk\ln\zeta = U + PV - TS$, by calculating from the 'sum over states' $\Sigma N_i/N = \quad e^{-\varepsilon 1/kT} + e^{-\varepsilon 2/kT} + e^{-\varepsilon 3/kT} + e^{-\varepsilon 4/kT} \ldots\ldots + e^{-\varepsilon n/kT} \ldots\ldots\ldots$; $nkT\ln\zeta$ is the thermodynamic potential (or free energy) for $n$ molecules, a function of an inversed action ratio $\zeta$. He defined the factor $1/\zeta$ as a function of the translational partition function $(2\pi mkT/h^2)^{3/2}V$, divided by the number of particles ($n$) – that is, as a translational action ratio as defined in this paper. By contrast its inverse $\zeta$ is an 'inaction' ratio indicating the free energy. For a perfect monatomic gas, $PV$ is equal to $RT$ and so $U + PV$ is equal to the enthalpy $H$, which does not change for individual molecules of a chemical species – unless the temperature changes.

In contrast to translation and rotation, vibrational action states higher than the ground state are largely unoccupied at ambient temperatures; most greenhouse molecules in the atmosphere are still in their coldest vibrational states, despite them radiating as required by the Stefan-Boltzmann equation proportional to the fourth power of the temperature in Kelvin. Thus vibrational action and entropy are minimal in Earth's atmosphere. This ordered state of low vibrational entropy is fortunate for life on Earth, otherwise stable molecules and structures would be impossible. In kinetic theory it has usually been assumed that molecular trajectories are linear, with no interaction between molecules. But the linear model is clearly only an approximation for translation. Whether the translational trajectory of the molecules is considered as curved or straight is irrelevant, given that the speed of energy transfer vastly exceeds that of the molecules; relatively to the thermal field energy bath referred to by Clausius in 1875, molecules are almost stationary.

It should be understood that all of the thermodynamic changes considered in this paper refer to changes possible in the physical state of small molecules in the atmosphere, mainly under conditions at the Earth's surface or using standard reference states. On Earth, only in the case of water do enthalpic changes associated with melting and vaporisation also need to be considered. Nor has attention been paid to changes in state following chemical reactions. In the case of chemical reactions the enthalpic changes related to internal action and entropy reflecting changes in bond energies either releasing or absorbing heat must also be included in the enthalpy as discussed above. However, chemical reactions will not be considered here.

Such reactions occurring at constant temperature require transfer of heat across the system's boundary, resulting in entropy changes external to the system. Action thermodynamics applied to transition states in chemical reactions will be considered in a separate paper. It is important to understand that chemistry at the Earth's surface is dependent on its physical environment. The phase changes discussed in this paper are strongly influenced by the thermodynamic and the gravitational properties of the milieu in which these changes occur.

***Greenhouse gases and temperature equilibration in the gravitational field***

The entropy calculations in this paper for gases have been considered for equilibrium conditions. This assumes that temperatures in the atmosphere equilibrate with surface insolation instantaneously. This is clearly not the case and the atmosphere will exhibit non-equilibrium conditions, constantly responding to the changing intensity of radiation from the Sun. This suggests that a key role of greenhouse gases may be their function in facilitating heat transfer by radiation and convection from the surface, cooling it and also in re-radiation from within the atmospheric profile, though consistent with its local temperature.

When atmospheric molecules are heated internally by absorption of infrared radiation, increasing their vibrational or rotational action and entropy, translational modes of action will immediately respond mediated through subsequent collisions. Relaxation times for such excited states are of the order of $10^{-7}$ sec. The absorption of a quantum will decrease their internal free energy whilst increasing their inertia and capacity to exert pressure, potentially doing gravitational work while moving to higher altitude and thus lowering the local temperature as kinetic energy declines. This may seem paradoxical but it is consistent with the virial theorem whereby increases in gravitational potential energy are matched by decreases in half its magnitude of kinetic energy (Kennedy, 2001, chapter 5 and Endnotes) and this idea will developed further in a companion paper on the barometric formula in assessing the possible role of greenhouse gases in global warming.

Such dissipation processes for absorption and emission of radiant energy may give a special role for greenhouse gases in the atmosphere, since the major gases nitrogen and oxygen have little if any such absorptive activity. Their presence enhances the rate of transfer of radiant energy from the Earth's surface to higher altitudes. Indeed, this could be the most important role of polyatomic gases like carbon dioxide. On the whole, greenhouse gases are regarded negatively because of their proposed role as agents in global warming; but it is important to also consider for experimental testing their possible benefits, such as elevating the atmosphere and cooling the surface of the Earth. Gases with higher heat capacities (including nitrogen and oxygen) also tend to cause the atmosphere to be more elevated, because the temperature lapse rate with altitude is less than for monatomic gases of similar mass. Thus an atmosphere of carbon dioxide of mass 44 Daltons would be more elevated than one of argon of mass 40, despite its greater weight.

We can examine the relative absorptivity of the greenhouse gases and the existence of absorptive-emissive lines in the infrared (Tables 3, 4), recalling that the earth's surface has a maximum emission range of around 5-30 μm wavelength (10000 $cm^{-1}$=1 μm, 1000 $cm^{-1}$=10 μm; 100 $cm^{-1}$ = 100 μm) whereas sunlight is confined to the 0.3-5 μm range. The longer wavelength of terrestrial radiation compared to sunlight is a result of absorption of sunlight by surface materials and re-equilibration of the quanta with the much cooler surface of the Earth, compared to the boiling ocean of hydrogen atoms of the Sun. Obviously, polyatomic molecules absorb in the 5-30 μm wavelength band of the infrared and the more complex the molecules are, the greater the number of absorptions.

The quanta associated with changes in rotational and translational action must be of longer wavelength, in the microwave and radiowave range of frequencies not resonant with the earth's major energy primary emissions from sunlight. But the infrared radiation absorbed by greenhouse molecules will be converted to these lower frequencies as a result of work done in subsequent molecular collisions during temperature equilibration in the atmosphere – the process known as equipartition. It is of interest that the quanta able to promote equality of kinetic temperature with equilibrated molecules range from infrared for vibrational freedom to microwaves for rotation and radiowaves for translation, thus broadening the spectrum of the energy involved.

The theory of quanta as enunciated by Planck (1913) pointed out that the intensity of black body radiation within an evacuated space equilibrated with matter bounding this space was entirely a function of temperature – despite this material particle being restricted to just one speck of carbon at the centre. Therefore, even a space containing mainly non-greenhouse gases such as nitrogen and oxygen should still radiate with the same intensity of infrared radiation, an essential feature for the correct operation of equipartition. Conversely, the quantum theory demands that very few (but not none) of the nitrogen and oxygen molecules will be excited vibrationally by such radiation, compared to greenhouse molecules. Furthermore, the intensity of the radiation passing through the boundary of a parcel of air is not a function of the proportion of the molecules able to be excited, or even their total number per unit volume, but only on the absolute temperature of the equilibrated surface. In a space relatively free of greenhouse gas molecules at a given temperature, the rays of infrared quanta will travel further before absorption so the total heat capacity for such quanta will be less. However, equilibrium assumes that the radiation density detected and radiated through the boundaries will always satisfy Stefan-Boltzmann theory for surfaces emitting radiation.

According to Clausius (1875) and the second law, to heat the Earth's surface as a net process the atmosphere would need to be hotter than the surface. Consistent with this principle, most of the temperature increase at the surface of the Earth from energy fed back from the atmosphere must be a result of the reversal of convective processes in high pressure zones, when air is descending. The fall of atmospheric gases from higher gravitational energy is a work process generating heat, all air molecules simultaneously gaining kinetic energy and radiating equivalent heat quanta as required by the virial theorem of Clausius. Clearly, this process can heat the surface, as occurs in high pressure zones or anticyclones. However, this transfer of heat from the atmosphere must be balanced by compensating transfers of radiant heat into the atmosphere in low pressure zones as gravitational work is performed using heat. These reversible processes demonstrate the Carnot principle that so impressed Clausius. Surprisingly, in climate science little attention is paid to the reversible transfers between heat and work that are implied in Lagrange's earlier identity relating the second derivative of the inertia of a system of particles ($I=\Sigma mr^2$) with respect to time and its kinetic ($T$) and potential energy ($V$); these can be considered as surrogates for heat and work in a gravitational or central force system.

$$\tfrac{1}{2}\mathrm{d}^2 I/\mathrm{d}t^2 = 2T + V \tag{11}$$

On the basis of reversibility so that the equality of equation (11) is zero, so $2T = -V$, inertial effects can be considered as sources of heating or cooling as seen in convection and advection near the Earth's surface (Kennedy, 2001, chapter 5). In fact, this equation could be considered as a basis for the whole of climatology, a contention we will explore elsewhere.

Table 4 also shows the specific frequencies of infrared radiation from the Earth that different gases such as $CO_2$, $N_2O$ and $CH_4$ will absorb. But a $CO_2$ molecule activated by IR-absorption to vibrate more vigorously will transfer most of this energy to other air molecules in the next collisions, thus increasing their action and entropy while dissipating the activated internal state and increasing their Gibbs energy. The relaxation time for such vibrational and rotational excitations is about $10^{-7}$ sec, given molecular collision rates of $10^9 – 10^{10}$ sec$^{-1}$, during which translational energies of all air are in increased by this dissipation. As such, quanta from the Earth's surface will not be exchanged between air molecules. Nevertheless, new infra-red and microwaves quanta will be generated appropriate for the reduced temperature with altitude.

Furthermore, the more dilute the gas (e.g. $N_2O$ and $CH_4$), the greater its translational entropy per molecule – although its vibrational and rotational entropies will be purely a function of temperature. Thus, on absorbing a specific quantum of IR-radiation (exciting molecular $^{\text{vibration}}$) such a dilute gas will have a larger disequilibrium between its vibrational action and its translational action. In a subsequent collision, the greater inertia and amplitude of the vibrating atom should cause a more efficient transfer of momentum to surrounding air molecules, irrespective of whether they are greenhouse gases or not.

So this thermodynamic force and heating effect is transferred to $N_2$ and $O_2$ as a result of collisions and the heated gases expand to higher altitude, exchanging their increased kinetic energy for increased gravitational energy and cooling as a result. Perhaps it is more apt to consider that the greenhouse gases such as water play an important role in holding up the sky, enabling reversible gravitational work, thereby cooling the atmosphere!

***Adiabatic processes***

The concept of reversible adiabatic changes in the atmosphere is often invoked in connection with parcels of air moving by convection or advection – with no heat entering or leaving the parcel. As a result, any changes in the thermodynamic state of the air parcel, such as temperature or pressure changes, must be achieved with the constraint of no transfer of heat into or out of the parcel. In effect, this requires that these changes be isoentropic, according to Clausius' definition of entropy (1875).

$$dS = \int dQ/T$$

As a result, given the case for the relationship between action and entropy made preceding, there must also be no change in the total relative action. Since translational action is given by $(3kTI_t)^{1/2}$, this requires that any adiabatic change in temperature $T$ must be accompanied by a compensating change in the inertia $I_t$. This is effectively achieved as a change in volume or density of the gas. For example, in a reversible Carnot cycle heat engine with a source temperature $T_{source}$ and a sink temperature of $T_{sink}$, a monatomic working fluid like helium or argon would obey the relationship

$(T_{sink}/T_{source})^{3/2} = V_2/V_3$

Here $V_2 \Rightarrow V_3$ represents an isoentropic process in which external work is done at the expense of the heat content of the working fluid, its temperature falling from $T_{source}$ to $T_{sink}$ and the volume increasing from $V_2$ to $V_3$. Some of the heat content of the parcel as measured by molecular temperature has apparently disappeared as the work of reversibly separating the molecules.

Adiabatic cooling during expansion of a working fluid while external work is being done is a feature of the Carnot cycle, with no heat flowing into or out of the engine. If no work is being done, such as expansion of a gas into a vacuum (previously prepared), no cooling will occur. Thus, an adiabatic expansion process *per se* is not the cause of the lowered temperature with altitude. The cooling effect called the adiabatic lapse rate for parcels of air expanding into a lower pressure zone higher in the atmosphere is a result of gravitational work being done. A subsequent paper will provide an alternative view of the lapse rate as a function of the virial theorem that increased gravitational potential of the atmosphere corresponds to a decrease in kinetic energy of half this magnitude.

We can write, for the isoentropic change per mole of monatomic gas, given the heat capacity at constant volume $C_v$ of a monatomic gas is $1.5R$ per $^oK$

$$\Delta S = R\ln(@_3/@_2)^3 = R\ln(3kT_3I_3/3kT_2I_2)^{3/2}$$

$$= R\ln(T_3r_3^2/T_2r_2^2)^{3/2} = C_v\ln(T_{sink}/T_{source}) + \mathrm{R}\ln(V_3/V_2) \quad (12)$$

Given $(T_{sink}/T_{source})^{3/2} = V_2/V_3$, the change in entropy must be zero. This follows since the effect of the decrease in temperature on action just balances the effect of the increase in volume. So if the increase in volume (or decrease in pressure at constant temperature) is 10 times greater, $(T_{sink}/T_{source})^{3/2}$ is equal to 0.1 and the ratio $T_{sink}/T_{source}$ is equal to 0.2154467.

In general, we can write that for Carnot cycles, $T_{sink}/T_{source}$ is equal to $(V_2/V_3)^{R/Cv}$ for working fluids of greater heat capacity. This means that diatomic gases like $N_2$ and $O_2$ would have a smaller temperature range for a 10-fold increase in volume, with $T_{sink}/T_{source}$ of 0.40. This also would translate to the atmosphere as a lower change in temperature with altitude than with the monatomic argon.

These adiabatic changes considered above occur at constant gravitational potential, as when a parcel of air moves laterally by advection, doing electrical work on a wind farm. When a parcel of air moves reversibly by adiabatic convection to a higher gravitational potential we have to consider the cooling effect of doing gravitational work in addition to changes in the atmospheric pressure. The lower the pressure exerted by the weight of the atmosphere above the parcel of air, the less pressure-volume work and heat is needed for expansion. However, almost the same amount of heat is required to raise the gravitational potential of air molecules no matter what the altitude. This variation in pressure-volume work brings into question the use of the dry adiabatic lapse rate (9.8 C/km) at all altitudes.

A descending parcel of air may be adiabatically compressed and spontaneously heats as gravitational potential declines causing kinetic work and internal heat-work varying free energy to be done on the air as it falls. We will show elsewhere that the increase in kinetic heat shown by the temperature increase at the expense of gravitational potential energy is matched by the decrease in Gibbs energy of the thermodynamic field, also consistent with the virial theorem. Furthermore, the capacity to do work of the air parcel declines as the atmospheric pressure increases and pressure-volume work becomes more costly.

As appropriate for statistical thermodynamics, these calculations of entropy and free energy relate only to the scale of randomized molecular motions of canonical ensembles. Neither the kinetic energy nor the “work-heat” or potential energy involved in convective and advective motions of parcels of air has been considered here. The thermal energy required to initiate these higher order motions (i.e. neither vibrational, rotational nor translational) is substantial, even though the kinetic energy generated is relatively minor compared to that of the randomized molecular motions. But the "work-heat" required for anticyclones and cyclones generated by thermal gradients in the gravitational field is substantial. The observation here that the major part of the heat required per molecule ($sT$) from absolute zero to 298 K relates to the "work-heat" compared to the sensible kinetic heat is a striking observation rarely made. For example, for argon the total entropy (18.6k per molecule or 154 J/C/mole), shown in Table 4 at 1 atm, is 12.4 times the increase in entropy from 0 K for kinetic motion alone (1.5$k$). At 0.01 atm in the atmosphere, the ratio is even greater. For all the molecules in a rotating parcel of air, the potential energy or "work-heat" of motion in these coherent though gaseous "fly-wheels" is orders of magnitude greater than the kinetic energy of their circulation. Even though the dissipation of this "work-heat" as frictional heat at higher latitudes is a major mechanism for the dispersal of solar energy from the equator towards the poles, this source of warming is rarely properly considered in climate models.

**Conclusion**

Unfortunately, in recent years thermodynamics and statistical mechanics have largely fallen into disuse, except by specialists. However, this should not continue, given the ease of calculation of entropy and of free energy of gases from action displayed in this paper. We will show elsewhere the utility of this approach. Action thermodynamics provides a realistic modeling approach available to all, easily applied mathematically and simplifying study of the links between heat and work and morphogenesis. We recommend more widespread application of this more explanatory approach that so aptly partitions the enthalpic and statistical aspects of entropy.

**Acknowledgements**

We wish to give credit to colleagues, including Barry Noller, Rodney Roughley, John Knight and many others. As thought alone, this work required no funding.

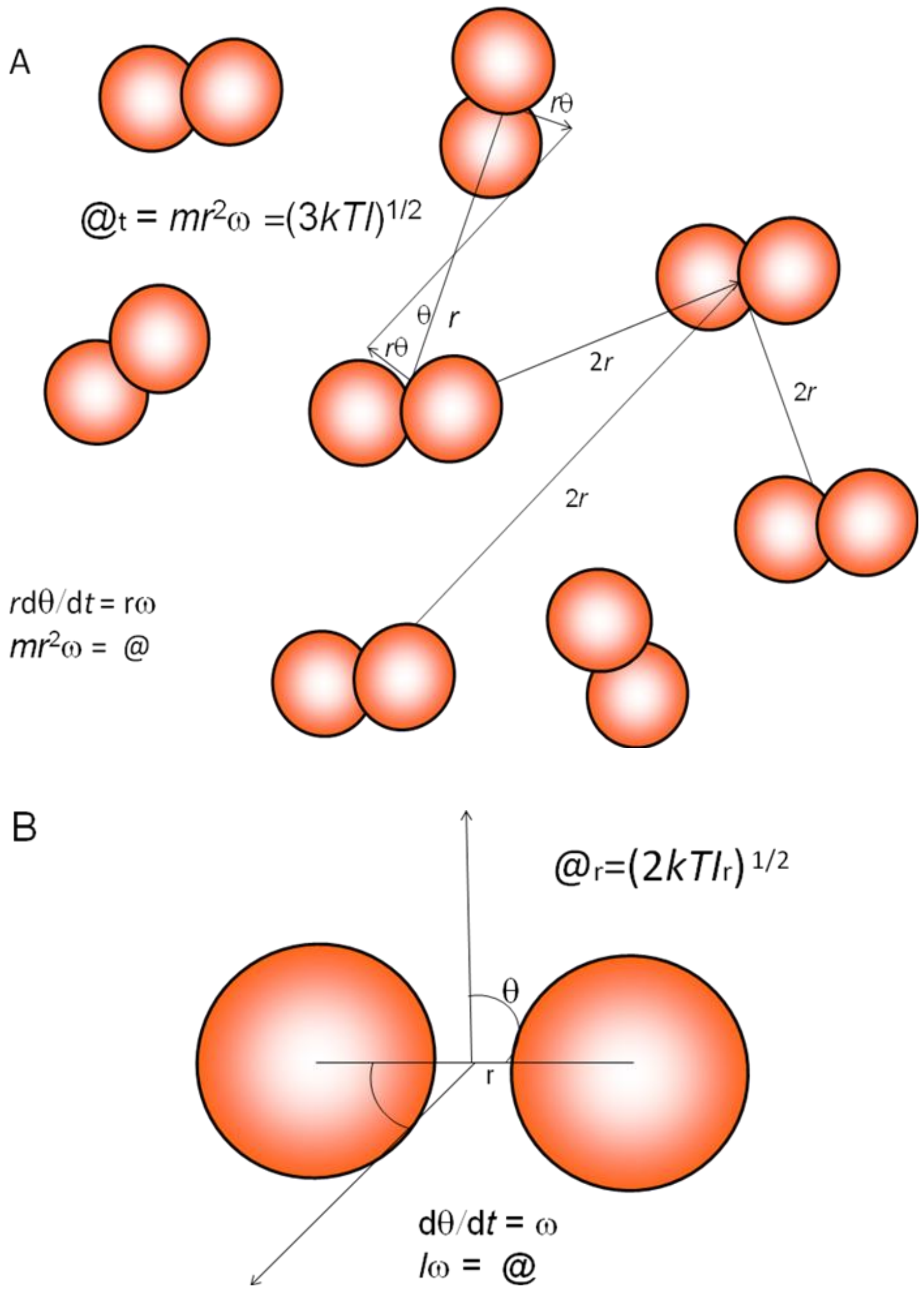

**Figure 1:** Calculation of translational and rotational action (@). Mean translational action $@_t$ (A) is estimated as explained in the text from average separation of $a = 2r$ by allocating each molecule space of $a^3 = V/N$ where $V$ is total volume and N is total number of diatomic molecules like dinitrogen ($N_2$). Relative angular motion $d\Theta/dt = \omega$ is estimated for molecules exhibiting the root-mean-square velocity, taking $3kT = mv^2 = mr^2\omega^2$. Then translational action is equal to $(3kTI_t)^{1/2}$. Rotational action $@_r$ (B) for linear molecules such as $N_2$, $O_2$ and $CO_2$ is estimated similarly, equated to $(2kTI_r)^{1/2}$.

.

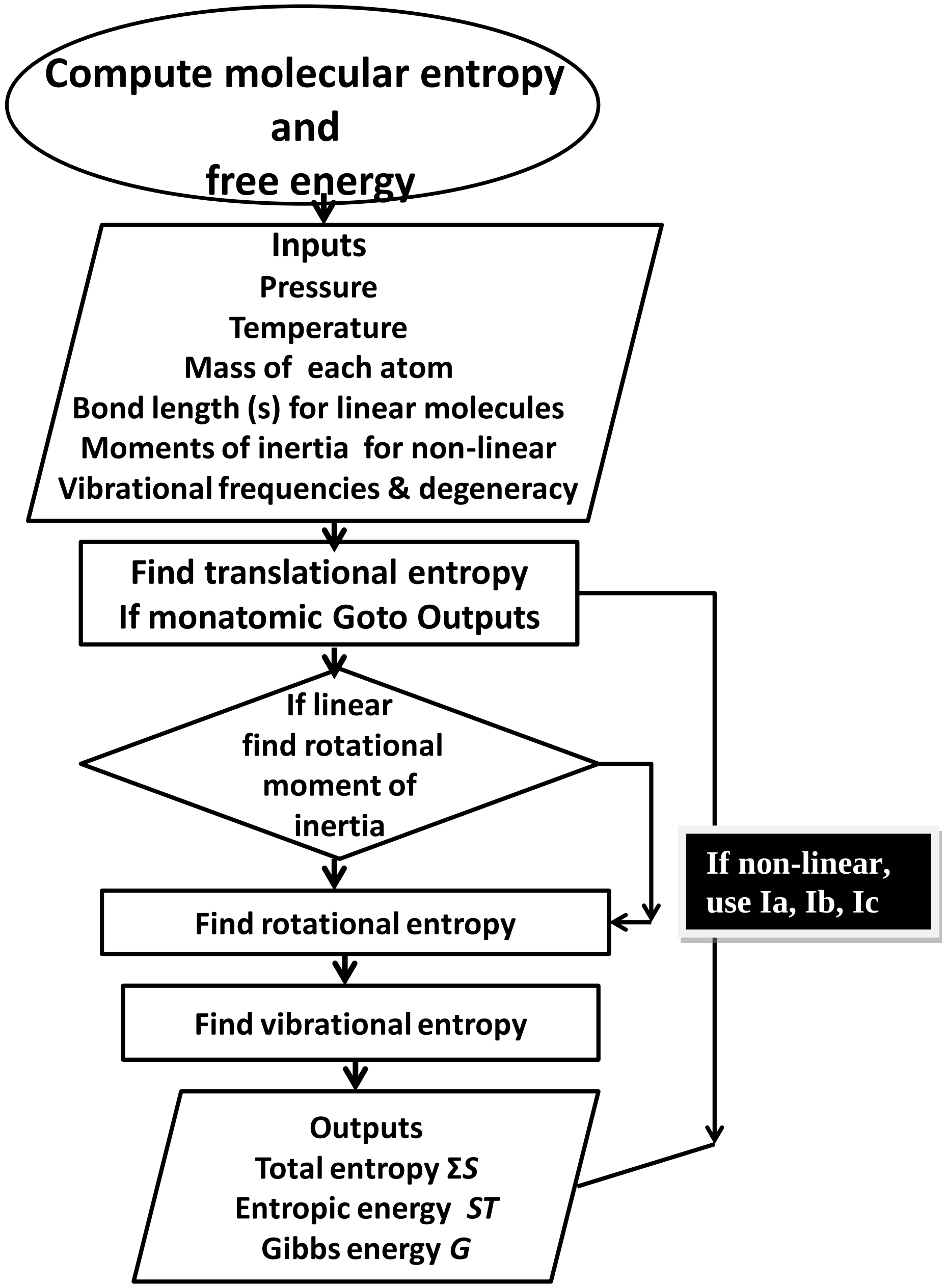

**Figure 2:** Flow diagram for computing absolute entropy and Gibbs energy. A fully annotated description of the relevant algorithms and subroutines to compute entropy and free energy is available on request to the corresponding author.